\documentclass[%
 reprint,
 superscriptaddress,
 numerical,
 showpacs,
 amsmath,amssymb,
 aps,
 longbibliography,
 prl,
 floatfix
]{revtex4-1}

\usepackage{graphicx}
\usepackage[usenames,dvipsnames,svgnames,table]{xcolor}
\usepackage[colorlinks=True,linkcolor=blue,citecolor=blue,urlcolor=blue]{hyperref}

\let\oldciteauthor=\citeauthor
\def\citeauthor#1{\hypersetup{citecolor=black}\oldciteauthor{#1}}

\let\oldciten=\onlinecite
\def\onlinecite#1{\hypersetup{citecolor=blue}\oldciten{#1}}

\let\oldcite=\cite
\def\cite#1{\hypersetup{citecolor=blue}\oldcite{#1}}

\makeatletter
\DeclareFontFamily{OMX}{MnSymbolE}{}
\DeclareSymbolFont{MnLargeSymbols}{OMX}{MnSymbolE}{m}{n}
\SetSymbolFont{MnLargeSymbols}{bold}{OMX}{MnSymbolE}{b}{n}
\DeclareFontShape{OMX}{MnSymbolE}{m}{n}{
    <-6>  MnSymbolE5
   <6-7>  MnSymbolE6
   <7-8>  MnSymbolE7
   <8-9>  MnSymbolE8
   <9-10> MnSymbolE9
  <10-12> MnSymbolE10
  <12->   MnSymbolE12
}{}
\DeclareFontShape{OMX}{MnSymbolE}{b}{n}{
    <-6>  MnSymbolE-Bold5
   <6-7>  MnSymbolE-Bold6
   <7-8>  MnSymbolE-Bold7
   <8-9>  MnSymbolE-Bold8
   <9-10> MnSymbolE-Bold9
  <10-12> MnSymbolE-Bold10
  <12->   MnSymbolE-Bold12
}{}

\let\llangle\@undefined
\let\rrangle\@undefined
\DeclareMathDelimiter{\llangle}{\mathopen}%
                     {MnLargeSymbols}{'164}{MnLargeSymbols}{'164}
\DeclareMathDelimiter{\rrangle}{\mathclose}%
                     {MnLargeSymbols}{'171}{MnLargeSymbols}{'171}
\makeatother

\newcommand{\ket}[1]{\left\vert #1 \right\rangle}
\newcommand{\bra}[1]{\left\langle #1 \right\vert}

\begin{document}

\title{Anomalous conductance scaling in strained Weyl semimetals}

\author{Jan Behrends}
\affiliation{Max-Planck-Institut f{\"u}r Physik komplexer Systeme, 01187 Dresden, Germany}
\affiliation{T.C.M. Group, Cavendish Laboratory, University of Cambridge, J.J. Thomson Avenue, Cambridge, CB3 0HE, United Kingdom}

\author{Roni Ilan}
\affiliation{Raymond and Beverly Sackler School of Physics and Astronomy, Tel-Aviv University, Tel-Aviv 69978, Israel}

\author{Jens H. Bardarson}
\affiliation{Department of Physics, KTH Royal Institute of Technology, Stockholm, SE-106 91 Sweden}

\begin{abstract}
Magnetotransport provides key experimental signatures in Weyl semimetals. 
The longitudinal magnetoresistance is linked to the chiral anomaly and the transversal magnetoresistance to the dominant charge relaxation mechanism.
Axial magnetic fields that act with opposite sign on opposite chiralities facilitate new transport experiments that probe the low-energy Weyl nodes.
As recently realized, these axial fields can be achieved by straining samples or adding inhomogeneities to them.
Here, we identify a robust signature of axial magnetic fields:
an anomalous scaling of the conductance in the diffusive ultraquantum regime.
In particular, we demonstrate that the longitudinal conductivity in the ultraquantum regime of a disordered Weyl semimetal subjected to an axial magnetic field increases with both the field strength and sample width due to a spatial separation of charge carriers.
We contrast axial magnetic with real magnetic fields to clearly distinguish the different behavior of the conductance.
Our results rely on numerical tight-binding simulations and are supported by analytical arguments.
We argue that the spatial separation of charge carriers can be used for directed currents in microstructured electronic devices.
\end{abstract}

\maketitle

Small variations in the parameters of a Hamiltonian describing a solid can affect the low-energy degrees of freedom in the form of a pseudofield, mimicking the effects of an external field.
Such pseudofields can have properties that are fundamentally different from the fields commonly realized in nature.
Inhomogeneities in Weyl semimetals~\cite{Weng:2015ec,Huang:2015ig,Xu:2015kb,Lv:2015fj}, for example, effectively induce~\footnote{Inhomogeneities additionally induce torsion and curvature to the underlying spacetime~\cite{Alvarez:1984ha,Parrikar:2014bv,Cortijo:2016hx,Weststrom:2017fy,Ferreiros:2019kz}, which we neglect.} axial pseudoelectric and pseudomagnetic fields~\cite{Chernodub:2014ez,Cortijo:2015ip,Pikulin:2016bn,Grushin:2016ji}, similar to strain-induced pseudofields in graphene~\cite{Manes:2007gc,Guinea:2010fl,Levy:2010hl}.
These axial fields act with opposite sign on left- and right-handed Weyl fermions, which constitute the low-energy degrees of freedom in Weyl semimetals, enriching their electrodynamics by a broader set of fields~\cite{Cortijo:2015ip,Landsteiner:2016kl,Behrends:2019dk,Ilan:2019io}.
As a consequence of axial magnetic fields, pseudo-Landau levels emerge at low energies~\cite{Pikulin:2016bn,Grushin:2016ji} while axial electric fields induce chirality-dependent transport~\cite{Gorbar:2017dn,Gorbar:2017ew}.
Axial magnetic fields are not unique to Weyl semimetals but may also be be simulated in cold atom systems~\cite{Roy:2018df}, and were demonstrated in various Weyl metamaterials~\cite{Peri:2019hn,Jia:2019dq}, but no definite manifestation has been found in the solid state.

Robust transport signatures for axial magnetic fields are still lacking.
Although predicted to enhance the conductivity of a Weyl semimetal by additional anomaly-induced contributions~\cite{Pikulin:2016bn,Grushin:2016ji}, akin to effects of external magnetic fields~\cite{Son:2013kd,Burkov:2015ei,Zhang:2016dx,Armitage:2018dg}, this enhancement relies on a local redistribution of charge~\cite{Behrends:2019dk}, which leads to screening effects that may spoil this contribution.
In this work, we identify an anomalous scaling of the conductance in the ultraquantum limit as a unique signature of axial magnetic fields.
This anomalous scaling, present for moderate disorder strengths, is induced by a spatial separation of counterpropagating modes, as we explain in the following.
This is fundamentally different from transport in the presence of a regular magnetic field, which in the ultraquantum limit is dominated by chiral modes that are separated in momentum space but overlapping in real space.

In the following, we focus on the simplest time-reversal breaking Weyl semimetals with only a single pair of Weyl nodes.
The momentum-space separation $2\mathbf{b}$ between the two Weyl nodes can at low energies be effectively regarded as a vector potential that acts with an opposite sign on these two Weyl nodes, a so-called axial vector potential.
Inhomogeneities, e.g., strain-induced variations in the hopping amplitudes, may result in a spatial variation, $\mathbf{b} \to \mathbf{b}(\mathbf{r})$~\cite{Pikulin:2016bn,Grushin:2016ji}.
Generally, a spatially varying $\mathbf{b}$ gives rise to an axial magnetic field via~\cite{Bertlmann:2000hn}
\begin{equation}
 \mathbf{B}_5 = \nabla \times \mathbf{b} .
\end{equation}
In a lattice, $\mathbf{b}$ must go to zero at the boundaries of an open system~\cite{Grushin:2016ji}; for periodic boundary condition, it must respect the periodicity~\cite{Behrends:2019dk}.
Thus, any axial magnetic field $\mathbf{B}_5$ averages to zero~\cite{Grushin:2016ji}, and any equilibrium current induced by $\mathbf{B}_5$ vanishes, when integrated over the whole sample~\cite{Sumiyoshi:2016bz,Kodama:2019bo}.

\begin{figure}
 \centering
 \includegraphics[width=\linewidth]{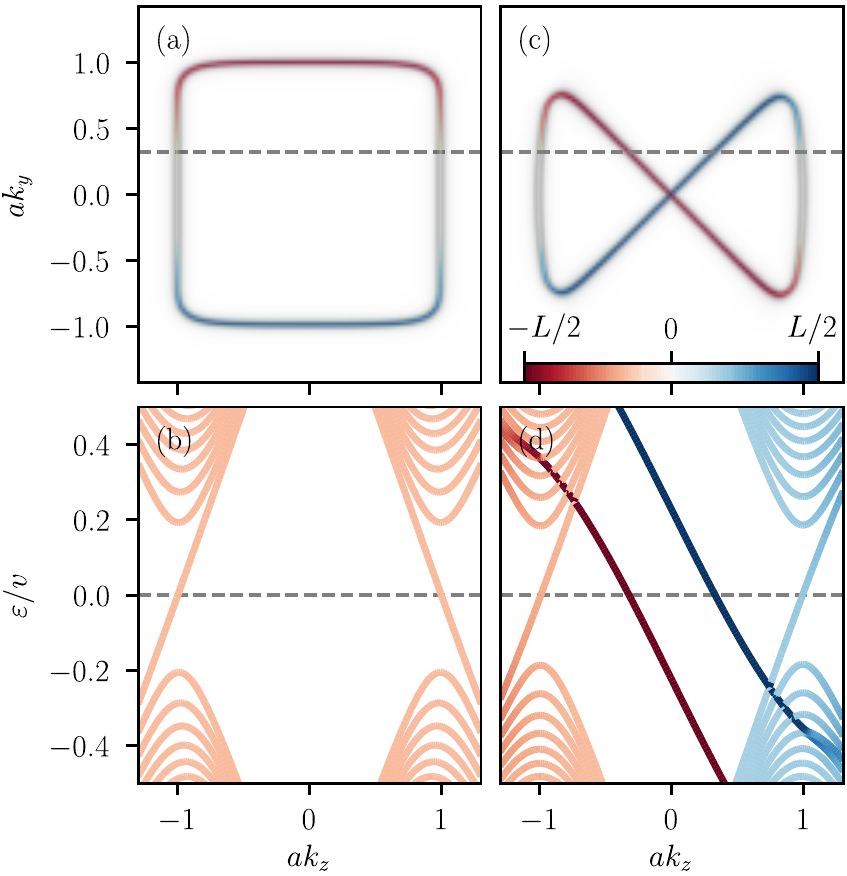}
 \caption{Fermi surface and dispersion of a Weyl semimetal, Eq.~\eqref{eq:ham_with_velocities}, subjected to a magnetic field $\mathbf{B}$, panels~(a) and~(b), and to an axial magnetic field $\mathbf{B}_5$, panels~(c) and~(d).
 The color denotes the real-space localization $\langle x \rangle$ of the states; cf. inset panel~(c).
 We choose open boundary conditions with $L/a=100$ sites in the $x$ direction and a (pseudo-) magnetic length of $\ell_B = \ell_5 = 7.07a$ with lattice constant $a$.
 (a) Fermi surface in presence of $\mathbf{B} = B \hat{z}$.
 (b) Corresponding dispersion at $k_y = 0.19/a$ [dashed lines in panels~(a) and~(c)].
 The dashed lines in panels (b) and (d) show the Fermi energy $\varepsilon_F =0$ as a guide for the eyes.
 (c) Fermi surface in presence of bulk $\mathbf{B}_5 = B_5 \hat{z}$ that results in pseudo-Landau levels with opposite position-momentum locking.
 (d) Corresponding dispersion at $k_y = 0.19/a$.
 }
 \label{fig:opposite_velocities}
\end{figure}

The consequences of axial magnetic fields on charge transport can be intuitively understood by comparing the dispersion in the presence of magnetic and axial magnetic fields for a pair of Weyl nodes separated by a constant component along $z$, shown in Fig.~\ref{fig:opposite_velocities}.
For concreteness we adopt a slab geometry and for the real magnetic field $\mathbf{B} = B\hat{z}$ we work in the Landau gauge $\mathbf{A} = Bx\hat{y}$, which leaves $k_y$ as a good quantum number.
In the presence of a magnetic field, the bulk Landau levels are exponentially localized at $x = \ell_B^2 k_y$ with localization length $\ell_B = \sqrt{\hbar/(e B)}$.
At each surface, Fermi arcs connect the bulk Landau levels of opposite chirality at the same momentum $k_y = L/(2 \ell_B^2)$, resulting in a rectangular Fermi surface [Fig.~\ref{fig:opposite_velocities}(a)].
Figure~\ref{fig:opposite_velocities}(b) shows the energy dispersion at a constant $k_y$.
The bulk zeroth Landau levels are counterpropagating, and localized at the same real-space position due to position-momentum locking.
Thus, backscattering is possible via scattering between these levels.

In the presence of a constant bulk $\mathbf{B}_5$, pseudo-Landau levels form.
Since $\mathbf{B}_5$ acts with opposite sign on the two chiralities, the bulk Landau levels propagate in the same direction.
Assuming the $\mathbf{B}_5$ is generated by a variation $\mathbf{b} \propto x\hat{y}$, these levels are exponentially localized at $x = \pm \ell_5^2 k_y$, with the sign determined by the chirality and the pseudo-magnetic length $\ell_5 = 1/\sqrt{B_5}$.
Surface states connect the bulk pseudo-Landau levels of opposite chiralities at different $k_y = \pm L/(2\ell_5^2)$, such that surface states twist in momentum space [Fig.~\ref{fig:opposite_velocities}(c)].
In this example, with a constant component $b_z \hat{z}$ parallel to $\mathbf{B}_5$, the Fermi surface traces a bowtie~\cite{Behrends:2019dk}.
A momentum-space-cut at constant $k_y$ [Fig.~\ref{fig:opposite_velocities}(d)] reveals that the bulk pseudo-Landau levels propagate in the same direction.
Thus, scattering between those states does not contribute to the transport time.
Since $\mathbf{B}_5$ averages to zero over the whole sample, surface states propagate in the opposite direction.
Accordingly, only scattering processes between the counterpropagating bulk and surface contribute to the transport time.
Charges localized deep in the bulk must scatter all the way to the surface to relax.

As a consequence of the limitation of backscattering to bulk-to-surface scattering, we predict a robust experimental signature of an axial magnetic field observable in simple transport measurements:
the scaling of conductance with the system's width.
We find that the conductance along $\mathbf{B}_5$ increases with the width cubed, different from the width squared, usually obtained in diffusive systems~\cite{AshcroftMermin}.
This anomalous scaling originates in the spatial separation of left- and right-moving modes, with one confined to the bulk and the other to the surface, as argued in Ref.~\onlinecite{Pikulin:2016bn}.
Because of the spatial separation, scattering mechanisms are substantially modified compared to standard magnetotransport experiments, as we elaborate on below. 
We explicitly demonstrate the scaling by employing tight-binding simulations, including the effects of disorder, to compute the conductance.
We use the Hamiltonian
\begin{align}
 \mathcal{H}_\mathbf{k} =& v \left[ \sin (k_y a) \sigma_x - \sin (k_x a) \sigma_y \right] \tau_z + v \sin (k_z a) \tau_y \nonumber \\
& + t \sum_i \left[ 1 - \cos (k_i a) \right] \tau_x + v a \mathbf{u} \cdot \mathbf{b} .
 \label{eq:ham_with_velocities}
\end{align}
with the lattice constant $a$, and $\mathbf{u} = ( -\sigma_x \tau_x, -\sigma_y \tau_x , \sigma_z )$.
The Pauli matrices $\sigma_\mu$ and $\tau_\mu$ act on different degrees of freedom, e.g., spin and orbital.
We discuss a time-reversal invariant model with four Weyl nodes, which allows for bulk-to-bulk backscattering, in the Appendix.
When $t = 2 v/\sqrt{3}$, as throughout this work, $2 \mathbf{b}$ equals the Weyl node separation in momentum space with deviations $O (\mathbf{b}^5)$ when $\mathbf{b}$ points along a primitive lattice vector, and deviations $O(\mathbf{b}^3)$ otherwise.
We choose $\mathbf{b} = (0,B_5 x,b_z)$ with $x \in [-L_x/2,L_x/2]$ to generate an axial magnetic field $\mathbf{B}_5 = B_5 \hat{z}$ parallel to the constant component $b_z$.
We take open boundary conditions in the $x$-direction, and average over twisted periodic boundary conditions in the $y$-direction $\psi (y+L_y) = e^{i\phi} \psi (y)$.
The transport direction is $z$, parallel to $\mathbf{B}_5$ and $\mathbf{B}$.

We perform all transport calculations using \textsc{Kwant}~\cite{Groth:2014ia}.
The total system consists of a sample of length $L_z = L_\parallel$ in the $z$ direction and width $L_x = L_y = L_\perp$ in both $x,y$ directions.
It is connected to two semi-infinite clean leads governed by the same Hamiltonian~\eqref{eq:ham_with_velocities} at the chemical potential $\mu_\mathrm{lead}$ with the same node separation and (axial) magnetic field.

For clean systems and $\mu_\mathrm{lead} = 0$, transport is ballistic, giving the conductance $G=n e^2/h$ for $n$ propagating modes.
For both $\mathbf{B}$ and $\mathbf{B}_5$, $n$ increases linearly with the sample's cross section $L_\perp^2$ and the field strength.
For $\mathbf{B}$, the number of propagating modes equals the degeneracy of the bulk Landau levels, and for $\mathbf{B}_5$, $n$ equals twice the degeneracy since surface states contribute equally to transport.
The Landau level degeneracy increases linearly with $L_\perp^2$ and $B$ ($B_5$).
Thus, for a clean system, the scaling of the conductance with $L_\perp^2$ and the field strength is the same for both $B$ and $B_5$.

\begin{figure}
 \includegraphics[width=\linewidth]{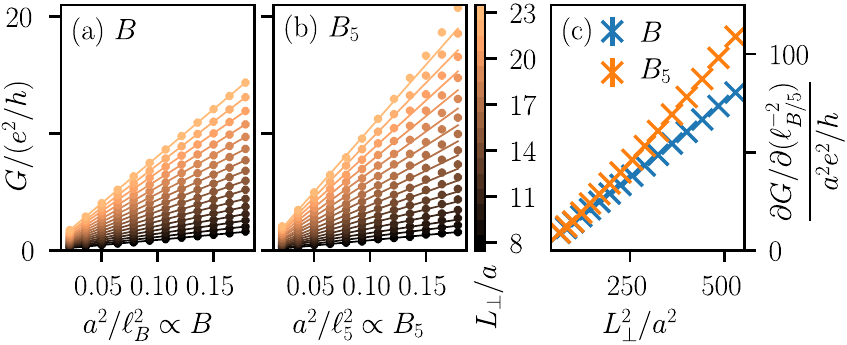}
 \caption{Conductance of clean samples in the presence of (axial) magnetic fields.
 (a)~Numerically obtained conductance (circles) as a function of the field strength $B$ ($\mathbf{B}_5=0$) for transport along $\hat{z} \parallel \mathbf{B}$.
 The conductance is independent of $L_\parallel$ and grows linearly with $B$ (linear fit indicated by solid lines).
 (b) Similarly, the conductance grows linearly with $B_5$ ($\mathbf{B} = 0$) up to  $a^2/\ell_5^2 \lesssim 0.12$.
 (c) The slope of $\partial G/\partial B_{(5)}$ increases linearly with $L_\perp^2$, such that $G \propto B L_\perp^2$ ($G \propto B_5 L_\perp^2$).
 The ratio of the slopes in panel (c) is $1.29$.
 }
 \label{fig:clean_conductance_BB5}
\end{figure}

In Figs.~\ref{fig:clean_conductance_BB5}(a) and~\ref{fig:clean_conductance_BB5}(b), we show the conductance as a function of $B$ and $B_5$, respectively, which indeed increases linearly with the field strength.
In the presence of $B_5$, the linear regime breaks down at a field strength characterized by $B_5 \propto a^2/\ell_5^2 \gtrsim 0.1$.
This breakdown is a lattice effect:
$\mathbf{b}$ equals the Weyl node separation only when neglecting higher-order corrections; the equality breaks down for large $| \mathbf{b} | \gg 1/a$, i.e., when $B_5 L_\perp \gg 1/a$.
Then, the conductance increases faster with $B_5$ [Fig.~\ref{fig:clean_conductance_BB5}(b)].
To avoid this lattice-induced effect, we limit the size of both $B_5$ and $L_\perp$ in the subsequent discussion.
Furthermore, we observe that the ratio of the conductance in the presence of $\mathbf{B}_5$ and $\mathbf{B}$ is smaller than $2$; cf.~Fig~\ref{fig:clean_conductance_BB5}(c).
This lattice effect only affects the total conductance in the presence of $B_5$, but not its scaling with the field strength or cross section, so we neglect it in our following analysis.

To discuss (axial) longitudinal magnetotransport in the presence of disorder, we add both vector and scalar disorder by introducing a random on-site matrix $V(\mathbf{r}_i)$ satisfying
\begin{align}
 & V (\mathbf{r}_i) = \sum_{\mu\nu} v_{\mu\nu} (\mathbf{r}_i) \sigma_\mu \tau_\nu, \\ 
 & \llangle v_{\mu\nu} (\mathbf{r}_i) v_{\rho\sigma} (\mathbf{r}_j) \rrangle =  g_{\mu\nu} K (\mathbf{r}_i- \mathbf{r}_j)\delta_{\mu\rho}\delta_{\nu\sigma}
 \label{eq:disorder}
\end{align}
to the lattice Hamiltonian.
The function $K (\mathbf{r}_i - \mathbf{r}_j)$ is the disorder correlator.
Since the presence of Weyl nodes in the Hamiltonian~\eqref{eq:ham_with_velocities} does not rely on any symmetries, no restrictions are put on the matrix structure of disorder and we set for convenience $g_{\mu\nu} =1$.

For a better comparison between real and axial magnetic fields, we briefly review previous work on the magnetoconductivity.
For systems larger than the mean free path, transport is diffusive and the conductance scales as $G =\sigma L_\perp^2/L_\parallel$ with a conductivity $\sigma$ independent of the system's dimensions~\cite{AshcroftMermin}.
The scaling of $\sigma$ with $B$ depends on the disorder type.
For white-noise disorder
\begin{equation}
 K (\mathbf{r}_i - \mathbf{r}_j) = \frac{W^2}{12} \delta_{ij} ,
 \label{eq:white_noise}
\end{equation}
the conductivity in the ultraquantum limit is independent of the magnetic field~\cite{Spivak:2016jy,Lu:2017gq}.
Gaussian correlations in the disorder potential with correlation length $\xi$ change the conductivity scaling to $\sigma \propto (\xi^2 + \ell_B^2)/\ell_B^2$~\cite{Lu:2015cj,Chen:2016gq,Zhang:2016hu}.
In the strong-field limit $\ell_B \ll \xi$, $\sigma$ increases with $\ell_B^{-2}$, i.e., with the magnetic field, as we explicitly demonstrate in the Appendix.
When $\ell_B \gg \xi$, the scaling for white-noise disorder is recovered, consistent with observations in TaAs~\cite{Ramshaw:2018dx}.

Replacing the magnetic field by an axial magnetic field results in two changes:
First, the conductivity increases linearly with $B_5$ for white-noise disorder and quadratically with $B_5$ for Gaussian-correlated disorder in the strong-field limit~\cite{Pikulin:2016bn}.
Second, as argued above, the conductivity increases with the system's width due to the spatial separation of counterpropagating modes.
As soon as disorder becomes large enough to mix the zeroth and higher Landau levels, all effects dominated by the zeroth Landau levels start to wash out---especially phenomena driven by the copropagating bulk zeroth Landau levels for $\mathbf{B}_5$ and counterpropagating Landau levels for $\mathbf{B}$.
We thus expect that the conductance scales similarly with field strength and system dimensions for axial and magnetic fields in strongly disordered systems.

To account for the scaling of the conductivity with the system's width, we define the dimensionless conductivity $g (L_\perp)$, a quantity \emph{not} independent of the system's dimensions, via
\begin{equation}
 G = \frac{e^2}{h a} \frac{L_\perp^2}{L_\parallel} g (L_\perp) .
 \label{eq:dimensionless_conductivity}
\end{equation}
We consider the regime where $g (L_\perp)$ is independent of $L_\parallel$, i.e., the diffusive regime where $G \propto 1/L_\parallel$.

To access the diffusive regime in tight-binding simulations, we need to avoid the ballistic and localized limits.
When $L_\parallel \gg L_\perp$, the system is essentially one dimensional and the charge carriers always localize, giving an exponentially decreasing conductance~\cite{Anderson:1958fz,Abrahams:1979iv}.
If the mean free path is larger than the system size, transport is ballistic, observable for $W/v \lesssim 1.5$ and the system sizes we consider.
To access the regime where the (axial) magnetic field influences transport, the (pseudo-) magnetic length [that sets the localization length of the (pseudo-) Landau levels] must be smaller than the sample thickness.
In our numerical simulations, the last restriction results in magnetic lengths ranging from $\ell_B = 2.6 a$ to $\ell_B = 4.8 a$, which require large (axial) magnetic fields; however, in mesoscopic samples, the condition $\ell_B /L \ll 1$ is much more easily satisfied than in the numerical simulations; i.e., only intermediate field strengths are necessary.

\begin{figure}
\centering
\includegraphics[width=\linewidth]{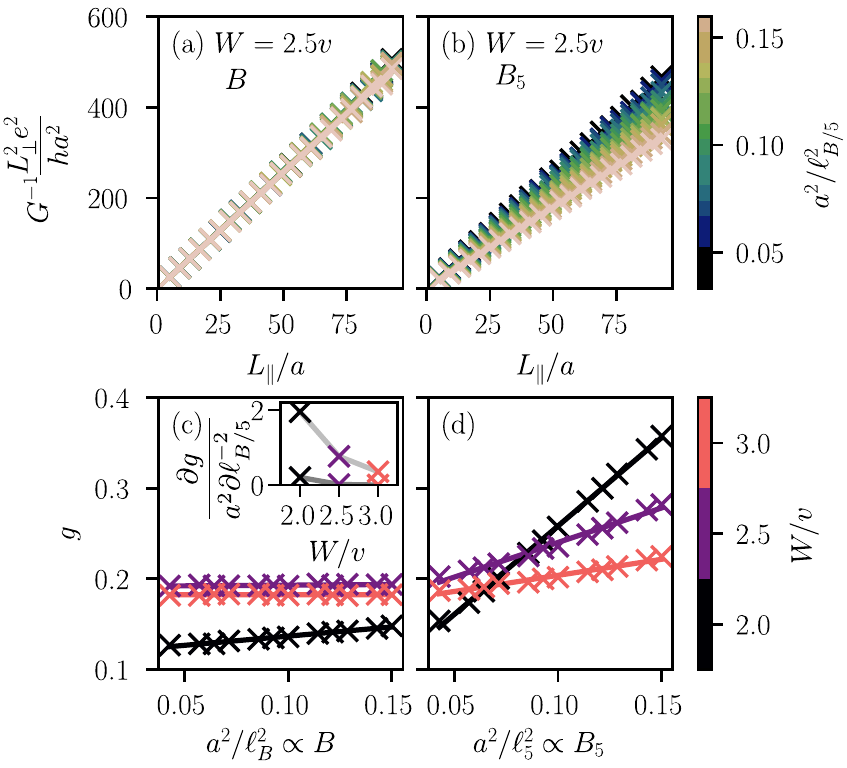}
\caption{Numerically obtained inverse conductance (a) in presence of $\mathbf{B}$ and (b) in presence of $\mathbf{B}_5$ as a function of $L_\parallel$, the system size in transport direction.
The colors denote different values of the (pseudo-) magnetic length $\ell_B$ ($\ell_5$).
The slope from panels (a) and (b) is the inverse of the dimensionless conductivity $g$, shown in panels (c) and (d) as a function of the (axial) magnetic field strength for different disorder strengths $W/v$.
While $g$ clearly increases with $B_5$, it stays almost constant in presence of $B$ and disorder $W/v \gtrsim 2.5$.
The inset in panel~(c) shows the slope $\partial g/\partial L_\perp$ for different disorder strengths and $B$ (dark gray) and $B_5$ (light gray).
All results are averaged over twisted boundary conditions and 100 disorder configurations; the chemical potential in the leads is $\mu_\mathrm{lead} = 1.5v$, and the transversal width $L_\perp=21a$.
}
\label{fig:conductivity_B}
\end{figure}

\begin{figure}
\centering
\includegraphics[width=\linewidth]{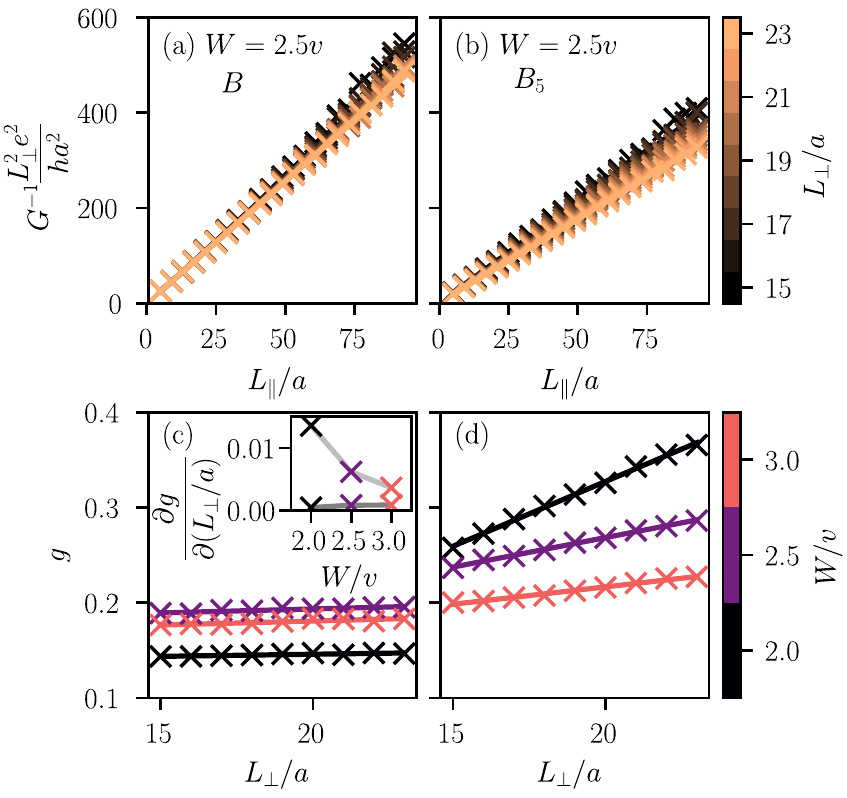}
\caption{Numerically obtained inverse conductance (a) in the presence of $\mathbf{B}$ and (b) in the presence of $\mathbf{B}_5$ as a function of $L_\parallel$.
The colors denote different values of $L_\perp$.
The slope from panels (a) and (b) is the inverse of the dimensionless conductivity $g$, shown in panels (c) and (d) as a function of $L_\perp$ for different disorder strengths $W/v$.
While $g$ increases linearly with $L_\perp$ in the presence of axial fields, it stays, apart from finite-size effects, constant in presence of magnetic fields $B$, as evident from the inset in panel (c) that shows the slope $\partial g/\partial L_\perp$ for $B$ (dark gray) and $B_5$ (light gray).
(Pseudo-) magnetic length $\ell_B = 2.65 a$ ($\ell_5 = 2.65 a$); other parameters as in Fig.~\ref{fig:conductivity_B}.
}
\label{fig:conductivity_L}
\end{figure}

We numerically investigate the conductance as a function of $L_\parallel$ in the presence of white-noise disorder~\eqref{eq:white_noise}.
We focus on the regime where the inverse conductance increases linearly with $L_\parallel$, which we can observe up to $L_\parallel \lesssim 3 L_\perp$ before reaching the one-dimensional limit showing onset of localization.
In Fig.~\ref{fig:conductivity_B}, we compare transport in the presence of axial fields with magnetic fields for different field strengths characterized by its (pseudo-) magnetic length.
Figures~\ref{fig:conductivity_B}(a) and~\ref{fig:conductivity_B}(b) show the conductance as a function of $L_\parallel$, normalized by the cross section $L_\perp^2$.
We find that the dimensionless conductivity~\eqref{eq:dimensionless_conductivity}, proportional to the inverse of the slope $\partial G^{-1}/\partial L_\parallel$, does not increase with $B$ for sufficiently large disorder $W \gtrsim 2.5v$ [Fig.~\ref{fig:conductivity_B}(c)] as expected from $\tau_v \propto 1/B$ for white-noise disorder~\cite{Lu:2015cj,Zhang:2016hu}.
In contrast, $g$ increases approximately linearly with the strength of the axial field $B_5$, Fig.~\ref{fig:conductivity_B}(d).
The slope $\partial g /\partial B_5$ decreases with the disorder strength $W$ before it eventually reaches zero; cf.\ inset in Fig.~\ref{fig:conductivity_B}(c).
When the disorder is larger than the spacing of the Landau levels ($W/v > a/\ell_B$ or $W/v > a/\ell_5$), higher Landau levels start to contribute to transport, allowing bulk-to-bulk backscattering, gradually decreasing the difference between $B$ and $B_5$ in the transport signature.

To contribute to transport relaxation, charges need to scatter the distance $L_\perp$, which takes $(L_\perp/\ell_5)^2 \propto B_5$ scattering events~\cite{Pikulin:2016bn}.
The scaling of the conductivity $g \propto B_5$ we observe is different from the prediction in Ref.~\onlinecite{Pikulin:2016bn}, since in our case $\tau_v \propto 1/B_5$ due to white-noise disorder.
For completeness, we analytically compute the scattering amplitude to lowest order in disorder in the Appendix and show that it decreases faster with system size than the amplitude from multiple scattering events.

We show the scaling of the conductivity as a function of $L_\perp$ in Fig.~\ref{fig:conductivity_L}.
As evident from Figs.~\ref{fig:conductivity_L}(a) and~\ref{fig:conductivity_L}(b), the inverse conductance $G^{-1}$ increases linearly with $L_\parallel$ when $L_\parallel \lesssim 3 L_\perp$ for both $B$ and $B_5$ and various values of $L_\perp$.
The dimensionless conductivity~\eqref{eq:dimensionless_conductivity} is independent of $L_\perp$ in the presence of $B$, Fig.~\ref{fig:conductivity_L}(c), but increases approximately linearly with $L_\perp$ in the presence of $B_5$, Fig.~\ref{fig:conductivity_L}(d).
This is one of our main results:
The conductivity increases linearly with the sample's thickness in the ultraquantum regime when $L_\perp \gg \ell_5$, corresponding to a conductance that increases with the thickness cubed, which we observe numerically for moderately large disorder ($1.5 \lesssim W/v \lesssim 3.5$).
The linear scaling of the conductivity with $L_\perp$ arises from subleading contributions in $L_\perp$ to the scattering time, as we show in the Appendix.
Similar to the scaling with $B_5$, the slope with $L_\perp$ decreases with the disorder strength, due to the mixing with higher Landau levels [inset in Fig.~\ref{fig:conductivity_L}(c)].
A dependence of the conductivity on a systems thickness is a known indicator for surface effects.
Indeed, in our case, the only effective transport relaxation is scattering between bulk and surface modes.
As the thickness grows, more modes localized deep in the bulk become less affected by such scattering, and the conductivity is enhanced. 

To conclude, in this work, we numerically investigated how charge transport in Weyl semimetals is affected by the presence of axial magnetic fields and white-noise disorder.
We explicitly demonstrated that the conductivity increases with both $B_5$, the strength of the axial field, and $L_\perp$, the transverse width of the system.
Both observations can be intuitively understood by the real-space separation of counterpropagating modes in the presence of $B_5$.

We are confident that the conductance scaling uncovered in this work can be observed in experiments, considering recent efforts in manufacturing microstructured Weyl semimetals~\cite{Bachmann:2017iz}, creating strain-induced fields in type-II Weyl semimetals~\cite{Kamboj:2019dd}, and transport experiments in Dirac nanowires~\cite{Li:2015bz,Wang:2017eo}.
As our results rely on time-reversal symmetry-breaking materials, magnetic-exchange-induced Weyl semimetals~\cite{Soh:2019cp,Su2019} are promising platforms for experiments.
The conditions necessary to observe the unusual scaling of the conductivity are sample width larger than the magnetic length, and sample length in the transport direction larger than the mean free path.
Our results show that the spatial separation of left- and right-moving charge carriers~\cite{Pikulin:2016bn} is stable towards small disorder, which can be used in devices that require a spatial separation of counterpropagating currents.
We stress that the implications for these observations are far reaching as they imply that experimentally approaching the ultraquantum limit of Weyl semimetals does not require the generation of large magnetic fields~\footnote{The ultraquantum limit requires large fields such that the chemical potential $\mu < \hbar v/\ell_B$, corresponding to fields $B>\mu^2/(\hbar e v^2)$. For a Fermi velocity $v=10^6\,\mathrm{m}/\mathrm{s}$ and $\mu = 50\,\mathrm{meV}$, fields of at least $B=4\,\mathrm{T}$ are necessary.}---moderate strain can easily induce axial field strengths of a few Tesla~\cite{Pikulin:2016bn,Grushin:2016ji,Liu:2017dn}.

\begin{acknowledgments}
We thank A.~G.~Grushin, D.~A.~Pesin, and D.~I.~Pikulin for insightful discussions.
This work was supported by the ERC Starting Grant No. 678795 TopInSy and the ERC Starting Grant No. 679722.
R.I. is supported by the ISF under Grant No. 1790/18.
\end{acknowledgments}

\appendix

\section{Appendix}

In this Appendix, we investigate transport for more lattice models and disorder types.
In particular, we show how axial fields influence transport in time-reversal symmetric Weyl semimetals, and how correlated disorder changes the conductance in a Weyl semimetal lattice model.

\subsection{Time reversal symmetric Weyl semimetals}

\begin{figure}
\includegraphics{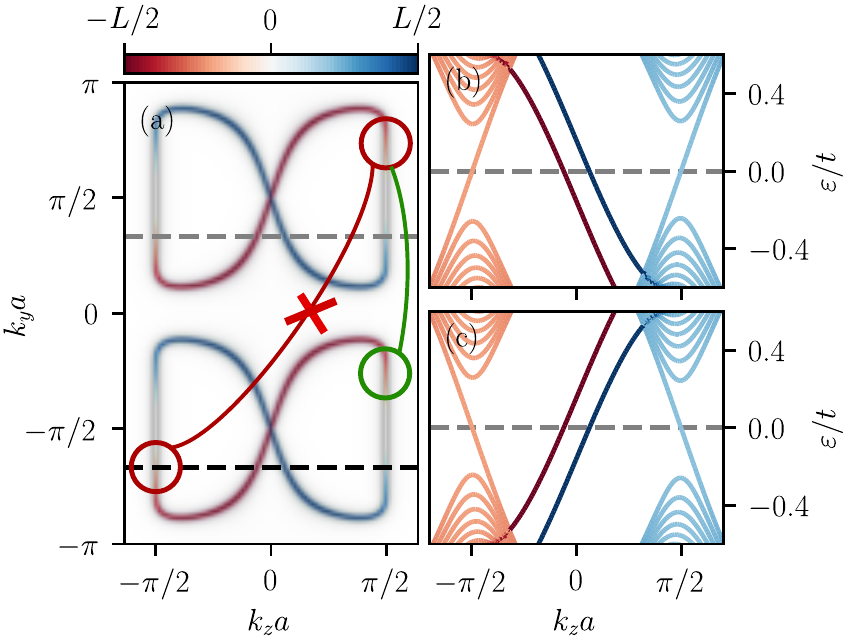}
\caption{(a) Fermi surface for the time-reversal invariant Hamiltonian~\eqref{eq:tr_invariant} in presence of an axial magnetic field $\mathbf{B}_5 = B_5 \hat{z}$ with open boundary conditions along $x$ with $L/a = 80$ lattice sites and the pseudo-magnetic length $\ell_5 = 5.61 a$.
The color denotes the real space localization of the wave functions, with red/blue corresponding to the two surfaces, cf.\ the small panel above~(a).
The lines red (green) lines that connect states (marked by circles) at the same real-space position denote backscattering processes forbidden (allowed) by time-reversal symmetry.
The black (gray) dashed line denotes a cut in momentum space at $k_y = -2.10/a$ ($k_y = 1.04/a$) with the dispersion along this cut shown in panel~(b) (panel~(c)).
(b) At $k_y = -2.10/a$, the bulk zeroth Landau levels are both right-moving, similar to the time-reversal model discussed in the main text.
(c) In contrast, at $k_y = 1.04/a$, the bulk zeroth Landau levels are both left-moving, which is imposed by time-reversal symmetry.
}
\label{fig:time_reversal}
\end{figure}

The model investigated in the main text explicitly breaks time-reversal symmetry by having a node separation $\mathbf{b}$ that corresponds to a magnetization.
Here, we employ a different model that hosts four Weyl nodes and does not break time-reversal symmetry, but only inversion symmetry.
In particular, we use the simple two-band model $\mathcal{H} =t \mathbf{d} \cdot \boldsymbol\sigma$ with the three components of $\mathbf{d}$
\begin{align}
 d_x =& \cos (b_x a) \sin (k_x a) + \sin (b_x a) \cos (k_x a) \sin ( k_z a) \nonumber \\
 d_y =& \cos (b_y a) \cos (k_y a) - \sin (b_y a) \sin (k_y a) \sin ( k_z a) \nonumber \\
 d_z =& 1 -\cos (k_z a) - \cos (b_x a) \cos (k_x a) \nonumber \\
	  & + \sin (b_x a) \sin (k_x a) \sin (k_z a)  .
 \label{eq:tr_invariant}
\end{align}
This model is a time-reversal symmetric extension of a commonly used two-band model~\cite{Yang:2011im} with the time-reversal operator $T= \sigma_z \mathcal{K}$, where $\mathcal{K}$ is complex conjugation.
It has two pairs of Weyl nodes: one pair at $k_{1,\chi} = \left( \chi b_x , \pi/2 + \chi b_y, -\chi \pi/2 \right)$ and another pair at $k_{2,\chi} = \left( -\chi b_x ,-\pi/2 -\chi b_y ,\chi \pi/2\right)$ with chirality $\chi = \pm 1$.
The first pair of Weyl nodes is separated by $\mathbf{b}$, and the other one by $-\mathbf{b}$, with $\mathbf{b} = (b_x,b_y,-\pi/2 )$.
An axial magnetic field induced by promoting $\mathbf{b}\to \mathbf{b}(\mathbf{r})$ conserves time-reversal symmetry.
We take $b_y \to B_5 x$ to obtain an axial magnetic field $\mathbf{B}_5 = B_5 \hat{z}$.
Fig.~\ref{fig:time_reversal} shows the resulting Fermi surface and dispersion.

To investigate magnetotransport, we introduce white-noise disorder similar to Eq.~\eqref{eq:disorder},
\begin{align}
 & V (\mathbf{r}_i) = \sum_\mu v_\mu (\mathbf{r}_i) \sigma_\mu \\
 & \llangle v_\mu (\mathbf{r}_i) v_\nu (\mathbf{r}_j) \rrangle
 = g_\mu K (\mathbf{r}_i - \mathbf{r}_j) \delta_{\mu\nu}
\end{align}
with $K (\mathbf{r}_i - \mathbf{r}_j) =\delta_{ij} W^2/12 $.
To obey time-reversal symmetry, the disorder must satisfy $T V (\mathbf{r}) T^{-1} = V (\mathbf{r})$, giving $g_\mu = 1 - \delta_{\mu,x}$, i.e, only the component $v_x (\mathbf{r}_i) =0$.
The presence of time-reversal symmetry excludes scattering between time-reversed partners, such that only one bulk zeroth Landau level is available for backscattering, as we illustrate in Fig.~\ref{fig:time_reversal}(a).
Different from the model with only two Weyl nodes, bulk-to-bulk backscattering is possible.
Thus, we expect contributions to the conductance from both bulk-to-bulk scattering and surface-to-surface scattering, the latter being especially important for the resistance at small system sizes [cf.\ Ref.~\cite{Gorbar:2016dy} for a discussion of the resistivity contribution of Fermi arcs].

\begin{figure}
 \includegraphics{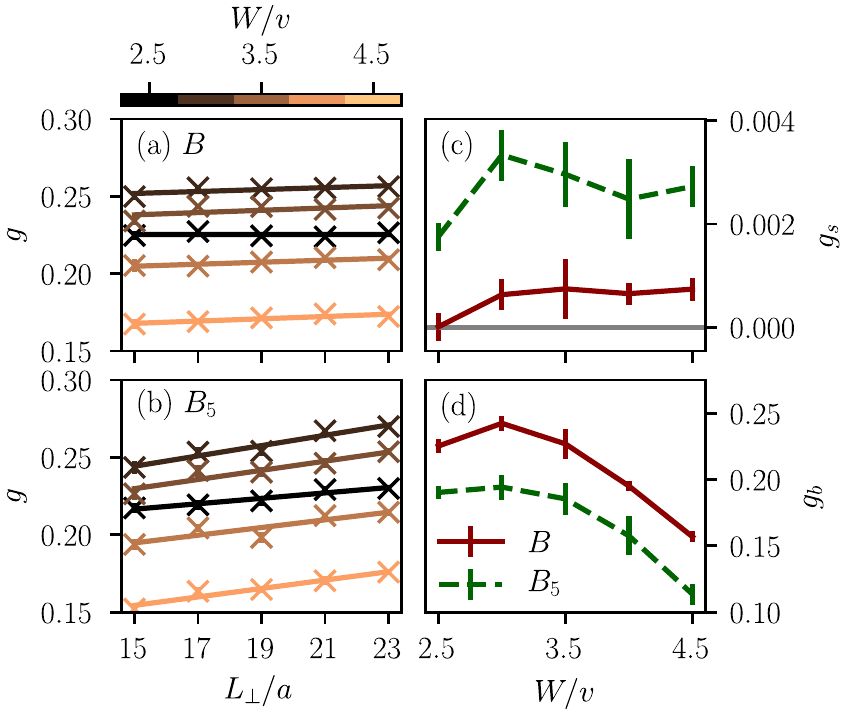}
 \caption{Scaling of the conductivity for time-reversal invariant Weyl semimetal with four Weyl nodes~\eqref{eq:tr_invariant}.
 (a) Dimensionless conductivity for a system subjected to a magnetic field characterized by the magnetic length $\ell_B = 2.96a$ and (b) subjected to an axial magnetic field with $\ell_5 = 2.96 a$.
 (c) From the data presented in panels (a) and (b), we obtain bulk and surface contributions separately via Eq.~\eqref{eq:bulk_surface}.
 }
 \label{fig:tr_invariant_vary_L}
\end{figure}

In Fig.~\ref{fig:tr_invariant_vary_L}(a) and~(b), we show the dimensionless conductivity $g$ for different transversal dimensions and disorder strengths for a disorder Weyl semimetal subjected to magnetic fields [panel~(a)] and axial magnetic fields [panel~(b)].
The dimensionless conductivity is defined in Eq.~\eqref{eq:dimensionless_conductivity}.
From the scaling with the transversal size $L_\perp$, we obtain the surface and bulk contributions from a linear fit
\begin{equation}
 g = g_s \frac{L_\perp}{a} + g_b .
 \label{eq:bulk_surface}
\end{equation}
In Fig.~\ref{fig:tr_invariant_vary_L}(c) and (d), we show $g_s$ and $g_b$ as a function of disorder strength for both $B$ and $B_5$.
The surface contribution $g_s$ is smaller for magnetic fields than for axial magnetic fields, as expected.
The bulk contribution $g_b$ scales similarly with disorder, and it is smaller for $B_5$ than for $B$.
The numerical results clarify that the anomalous scaling of the conductance in presence of $B_5$ does not hold for time-reversal symmetric models.

\subsection{Comparison of correlated and white-noise disorder}

In the main body of this work, we have only focused on white noise disorder.
In the presence of correlated disorder, it is not only the real-space position of counterprogating modes that determines the transport scattering time, but also their momentum-space position.
For physically relevant disorder correlations, e.g., Gaussian correlated disorder, scattering between states decreases with their momentum space separation, giving a internode scattering rate (scattering between different chiralities) that is much bigger than the intranode scattering rate (scattering within a chirality)~\cite{Spivak:2016jy,Zhang:2016hu,Lu:2017gq}.
In Weyl semimetals, this effect contributes to the large chiral-anomaly induced contribution to the magnetoresistance~\cite{Son:2013kd,Xiong:2015kl,Huang:2015gy,Shen:2016bp,Li:2016dp,Li:2016bj}, and its anisotropy~\cite{Arnold:2016hp,Reis:2016iq,Xiong:2016bx}.

For disorder with Gaussian correlations
\begin{equation}
 K (\mathbf{r}_i -\mathbf{r}_j) 
 = \frac{1}{\sqrt{2\pi\xi^2}} \frac{W^2}{12} \exp \left[ -\frac{(\mathbf{r}_i - \mathbf{r}_j)^2}{2 \xi^2} \right] ,
\end{equation}
the internode scattering rate scales as
\begin{equation}
 \frac{1}{\tau_v} \propto \frac{\xi^2}{\ell_B^2 + \xi^2}
 = \begin{cases} 1 & \ell_B \ll \xi \\ \xi^2/\ell_B^2 & \ell_B \gg \xi .
 \end{cases}
\end{equation}
Using $g \propto B \tau_v$, which holds in the ultraquantum limit and for weak disorder, we obtain the $g \propto 1$ for weak fields ($\ell_B \gg \xi$) and $g \propto B$ for strong fields ($\ell_B \ll \xi$).

\begin{figure}
 \includegraphics[width=\linewidth]{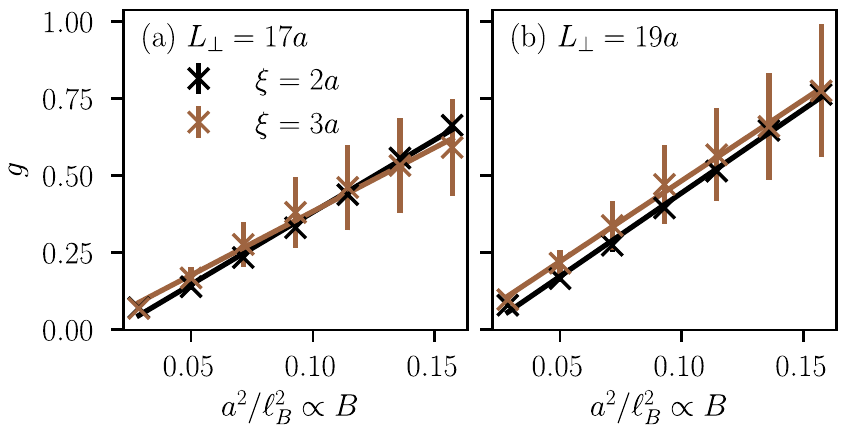}
 \caption{Dimensionless conductivity $g$ for two different disorder correlation lengths $\xi$ as a function of magnetic field.
 Different from white-noise disorder, the conductivity increases linearly with the applied magnetic field.
 We show data for two different transversal system sizes (a) $L_\perp = 17a$, and (b) $L_\perp = 19a$.
 In the limit $L_\perp \gg a$, the conductivity is independent of the transversal system size.
 The magnetic length ranges from $\ell_B = 2.52 a$ to $\ell_B = 5.92 a$ and the disorder strength is $W=7v$.
 }
 \label{fig:correlated_disorder}
\end{figure}

In Fig.~\ref{fig:correlated_disorder}, we show the dimensionless conductivity evaluated for a Weyl semimetal lattice model, Eq.~\eqref{eq:ham_with_velocities}, subjected to correlated disorder characterized by the correlation length $\xi$, and a longitudinal magnetic field.
We show that the conductivity increases linearly with the applied field, as expected in this regime.

\subsection{Perturbative analytical analysis of scattering amplitudes}

In this section, we investigate which mechanism may explain the observed scaling of the conductance in presence of an axial magnetic field.
We consider two different possibilities for backscattering, i.e., for bulk-to-surface scattering:
\begin{enumerate}
 \item Direct scattering, where we consider only the lowest order expansion in perturbation theory, the Born approximation.
 \item Multiple scattering events.
\end{enumerate}
It turns out direct scattering decreases exponentially with the transversal system size, while higher order scattering events only decrease as a power law when $L_\perp \gg \ell_{5}$.

\begin{figure}
 \includegraphics{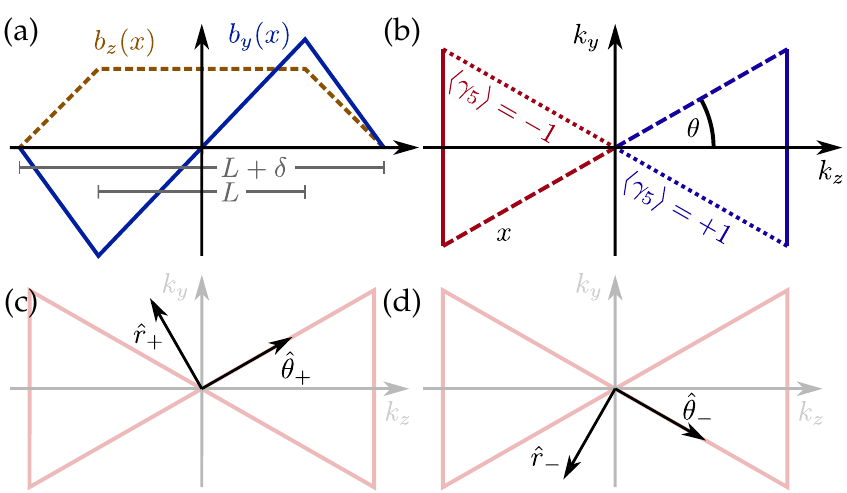}
 \caption{Setup for the system considered for the analytical calculation.
 (a) Position-dependence of the components of~$\mathbf{b}$.
 In the bulk, when $x \in [-L/2,L/2]$, the node separation has a constant component along $\hat{z}$, and a spatially varying component that results in $\mathbf{B}_{5} = \mathrm{const.}$
 At the surfaces of width $\delta/2$, both components are position dependent to ensure that $\mathbf{b}$ goes to zero in the vacuum.
 (b) Resulting bow-tie Fermi surface.
 Dotted/dashed lines denote states localized in the surface regions, solid lines denotes states in the bulk.
 The chirality changes, in the approximation we use, instantly from right-handed to left-handed at $k_z=0$, as denoted by the color code.
 This instant change is different from a lattice model, where a finite mass at the boundary couples both chiralities, resulting in a smooth transition.
 (c), (d) Definitions of the basis vectors $\hat{r}_{\sigma}$ and $\hat{\theta}_{\sigma}$ that we use in the analytical calculation.
 }
 \label{fig:setup}
\end{figure}

To obtain the direct bulk-to-surface scattering amplitude, we use the low-energy Hamiltonian of chirality~$\chi$
\begin{equation}
 \mathcal{H}^\chi = \chi \hbar v \left( \mathbf{k} - \chi \mathbf{b} \right)\cdot \boldsymbol\sigma .
\end{equation}
We consider a system comprised of a bulk and two surface regions.
In the bulk, $x\in [-L/2,L/2]$, the node separation has a constant component $\mathbf{b}_0 = b_0 \hat{z}$ and a spatially varying component that generates $\mathbf{B}_5 = B_5 \hat{z}$.
To counteract the bulk $B_5$, the node separation must go to zero at the surfaces.
We add two surface regions of width $\delta/2$ on each side of the sample.
As the node separation needs to go to zero at the surface, we choose
\begin{equation}
 \mathbf{b} (x) =
 \begin{cases}
 \frac{L+\delta +2 x}{\delta} \left( -\frac{B_5 L}{2} \hat{y} + b_0 \hat{z} \right)
		& -\frac{L+\delta}{2} < x \le -\frac{L}{2} \\
  B_5 x \hat{y} + b_0 \hat{z}
  		& -\frac{L}{2} < x < \frac{L}{2} \\
 \frac{L+\delta -2 x}{\delta} \left(  \frac{B_5 L}{2} \hat{y} + b_0 \hat{z} \right) & \frac{L}{2} \le x < \frac{L+\delta}{2}.
 \end{cases}
\end{equation}
The resulting Fermi surface resembles the bow tie discussed in the main text, but it not smooth due to the sharp corners [Fig.~\ref{fig:setup}].

Similar to the bulk Hamiltonian, the surface Hamiltonian has a nonzero axial field.
We label the two surfaces by $\sigma = \pm 1$ and introduce the rotated vectors,
\begin{align}
 \hat{r}_{\sigma}      =  \sigma \sin \theta \hat{y} - \cos \theta \hat{z} , & &
 \hat{\theta}_{\sigma} =  \sigma \cos \theta \hat{y} + \sin \theta \hat{z} ,  \nonumber\\
 \cos \theta = \frac{B_5 L}{\sqrt{ (B_5 L)^2 + 4 b_0^2}} , & &
 \sin \theta = \frac{2 b_0}{\sqrt{ (B_5 L)^2 + 4 b_0^2}} ,
\end{align}
such that the node separation at the surface reads
\begin{equation}
 \mathbf{b}_\sigma = \left( \frac{L+\delta}{2\,{\ell_5'}^2} - \frac{x}{{\ell_5'}^2} \right) \hat{\theta}_{\sigma}
\end{equation}
with the pseudo-magnetic length at the surface
\begin{equation}
 \frac{1}{{\ell_5'}^2} = \frac{1}{\delta} \sqrt{ 4 b_0^2 + B_5^2 L^2} .
\end{equation}
The resulting axial field is
\begin{align}
 \mathbf{B}_{5,\sigma}
 &= \nabla \times \mathbf{b} =
 \sigma \frac{2 b_0}{\delta} \hat{y}  - \frac{L}{\delta} B_5 \hat{z} = \frac{1}{{\ell_5'}^2} \hat{r}_{\sigma} .
\end{align}
To have a more convenient basis in terms of Landau levels, we rotate the surface Hamiltonian $\mathcal{H}^\chi_\sigma$ of chirality $\chi$ at the surface $\sigma$ by a unitary transformation
\begin{equation}
 {\mathcal{H}^\chi_\sigma}' =  \mathcal{U}_\sigma^\dagger \mathcal{H}^\chi_\sigma \mathcal{U}_\sigma ,
\end{equation}
giving
\begin{align}
 {\mathcal{H}^{(+)}_\sigma}' &= \frac{\hbar v}{\ell_5'}
 \begin{pmatrix}
  \ell_5' k_{r,\sigma} & -\sqrt{2} i a_{+,\sigma}^\dagger \\
  \sqrt{2} i a_{+,\sigma} & -\ell_5' k_{r,\sigma}
 \end{pmatrix} \\
 {\mathcal{H}^{(-)}_\sigma}' &= \frac{\hbar v}{\ell_5'}
 \begin{pmatrix}
  \ell_5' k_{r,\sigma} & -\sqrt{2} i a_{+,\sigma} \\
  \sqrt{2} i a_{+,\sigma}^\dagger & -\ell_5' k_{r,\sigma}
 \end{pmatrix}
\end{align}
with $k_{r,\sigma} = \mathbf{k} \cdot \hat{r}_{\sigma}$, $k_{\theta,\sigma} = \mathbf{k} \cdot \hat{\theta}_{\sigma}$, and
\begin{equation}
 a_{\chi,\sigma} = \frac{1}{\sqrt{2}} \left( \frac{x}{\ell_5'} - \sigma \left( \frac{L+\delta}{2 \ell_5'} - \chi \ell_5' k_{\theta,\sigma}\right) + i \ell_5' k_x \right) .
\end{equation}

\begin{figure}
 \includegraphics[width=\linewidth]{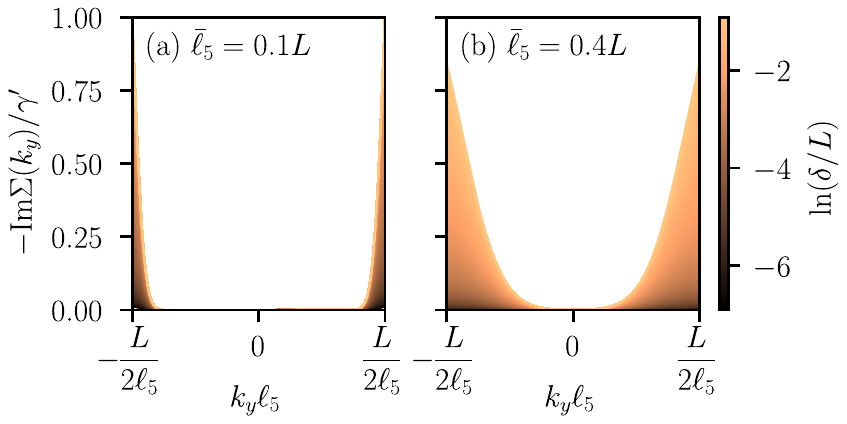}
 \caption{Analytical solution of the imaginary part of the self-energy correction, which is proportional to the inverse transport time $\tau_v^{-1}$.
 We rescale the self-energy correction by $\gamma' = \gamma^2 / (\pi \hbar v {\ell_5'}^2)$ and plot two different values of $\bar{\ell}_5 = \sqrt{\ell_5^2 + {\ell_5'}^2}$ in panels (a) and (b).
 The self-energy correction depends on $k_y$, which corresponds to the real space position $x = \ell_5^2 k_y$.
 Deep in the bulk, the transport time goes to infinity, since states cannot backscatter.
 Close to the surface, the transport time increases, since backscattering is possible.}
 \label{fig:analytical_tauv}
\end{figure}

As argued in the main text, backscattering is the only relevant mechanism for transport.
Since the pseudo-Landau levels in the bulk are chiral, the only way to scatter in the opposite direction is via relaxation through the surfaces.
The same goes for transport via the surface states---they need to scatter to the bulk of the system to relax.
We consider the self-energy that describes backscattering for both surfaces and both chiralities
\begin{equation}
 \hat{\Sigma}^\chi = \sum_{\chi',\sigma} \hat{V} \mathcal{G}^{\chi'\sigma} \hat{V} ,
\end{equation}
or, in Landau level basis
\begin{align}
 \Sigma_m^\chi (\omega,\mathbf{k}_\parallel)
 =& \sum_{\chi',\sigma}\sum_{m',\mathbf{k}_\parallel'} \bra{\Phi_{m,\mathbf{k}_\parallel}^\chi} \hat{V} \mathcal{U}_\sigma \ket{\Phi_{m',\mathbf{k}_\parallel'}^{\chi'\sigma}} \mathcal{G}_{m'}^{\chi'\sigma} (\omega,\mathbf{k}_\parallel' ) \nonumber \\
 & \times \bra{\Phi_{m',\mathbf{k}_\parallel'}^{\chi'\sigma}} \mathcal{U}_\sigma^\dagger \hat{V} \ket{\Phi_{m,\mathbf{k}_\parallel}^\chi} ,
\end{align}
where states with an additional label $\sigma$ are located at the surface and states without the $\sigma$ label are located in the bulk.
Similar to the approach in Ref.~\onlinecite{Behrends:2017fh}, we introduce a disorder correlator in Landau level basis, $\Gamma_{mm'}$, to write
\begin{equation}
 \Sigma_m^\chi (\omega,\mathbf{k}_\parallel)
 = \sum_{\chi',\sigma}\sum_{m',\mathbf{k}_\parallel'} \Gamma_{mm'}^{\chi\chi'\sigma} \mathcal{G}_{m'}^{\chi'\sigma} (\omega,\mathbf{k}_\parallel' ) .
\end{equation}
The form of $\Gamma_{mm'}$ naturally depends on the form of the disorder correlator in momentum space.
There are no symmetry restrictions on disorder, which means that all Pauli matrices are allowed in the disorder operator $\hat{V} = \sum_\mu \hat{v}_\mu \sigma_\mu$ with the correlations
\begin{equation}
 \llangle v_\mu (\mathbf{q})  v_\nu (-\mathbf{q}') \rrangle = K_\mu (\mathbf{q}) \delta_{\mu\nu} \delta_{\mathbf{q},\mathbf{q}'} .
\end{equation}
After averaging over disorder, the correlator becomes
\begin{align}
 \Gamma_{mm'}^{\chi\chi'\sigma} =& \sum_{q_x,\mu} K_\mu(\mathbf{q}) \int d x \Phi_{m \mathbf{k}_\parallel}^{\chi\,\dagger} (x) \sigma_\mu \mathcal{U}_\sigma \Phi_{m'\mathbf{k}_\parallel'}^{\chi'\sigma} (x)e^{i q_x x} \nonumber \\
 & \times \int d x' \Phi_{m'\mathbf{k}_\parallel'}^{\chi'\sigma\,\dagger} (x') \mathcal{U}_\sigma^\dagger \sigma_\mu \Phi_{m \mathbf{k}_\parallel}^{\chi} (x') e^{-i q_x x'}
 \label{eq:disorder_correlator}
\end{align}
For simplicity, we restrict to the zeroth Landau level.
The spinors are then
\begin{align}
 \Phi_{0\mathbf{k}_\parallel}^{(+)} (x)
	= ( \phi_{0,k_y}^{(+)} (x), 0 ) , & &
 \Phi_{0\mathbf{k}_\parallel}^{(-)} (x)
	= ( 0, \phi_{0,k_y}^{(-)} (x) ) \nonumber \\
 \Phi_{0\mathbf{k}_\parallel}^{(+)\sigma} (x)
 	= ( \phi_{0,k_{\theta,\sigma}}^{(+)\sigma} (x) ,0 ), & & 
 \Phi_{0\mathbf{k}_\parallel}^{(-)\sigma} (x)
	= ( 0, \phi_{0,k_{\theta,\sigma}}^{(-)\sigma} (x) ) .
\end{align}
Inserting these states into Eq.~\eqref{eq:disorder_correlator} yields
\begin{equation}
 \Gamma_{00}^{\chi\chi'\sigma} = \sum_\mu \frac{1 - g_\mu \chi \chi' \cos \theta}{2 \ell_5 \ell_5'} \sum_{q_x} K_\mu (\mathbf{q}) \mathcal{I}_+ \mathcal{I}_-
\end{equation}
with $g_\mu = (1,-1,-1,1)$ and the integral
\begin{align}
\mathcal{I}_\pm
  &= \int d x  \psi_{0} \left( \tfrac{x}{\ell_5} - \alpha \right) \psi_{0} \left( \tfrac{x}{\ell_5'} - \alpha' \right) e^{\pm i q_x x} \nonumber \\
  &= \frac{\sqrt{2} \ell_5  \ell_5'}{\sqrt{\ell_5^2+{\ell_5'}^2}}  e^{ - \frac{q_x^2 \ell_5^2 {\ell_5'}^2 \mp 2 i q_x \ell_5 {\ell'_5} ( \alpha' \ell_5 +\alpha \ell_5') + (\alpha \ell_5 - \alpha'\ell_5')^2}{2 \left(\ell_5 ^2+{\ell_5'}^2\right)} } ,
\end{align}
with the definition
\begin{align}
 \alpha = \chi \ell_5 k_y, & &  \alpha' = \sigma( \tfrac{ L + \delta}{2\ell_5'} -  \chi'\ell_5' k_{\theta,\sigma}' ) .
\end{align}
This simplifies the expression for the disorder correlator in Landau level basis
\begin{align}
 \Gamma_{00}^{\chi\chi'\sigma}
 =& \sum_\mu  \frac{(1 - g_\mu \chi \chi' \cos \theta) \ell_5 \ell_5'}{\ell_5^2+{\ell_5'}^2}  \\
 & \times \sum_{q_x} K_\mu (\mathbf{q}) e^{ - \frac{q_x^2 \ell_5^2 {\ell_5'}^2 + (\alpha \ell_5 - \alpha'\ell_5')^2}{\ell_5^2+{\ell_5'}^2} } . \nonumber 
\end{align}
To obtain the imaginary part of the self-energy, $\Gamma$ needs to be integrated over momentum
\begin{equation}
  \mathrm{Im} \Sigma_{00}^{\chi\chi'\sigma} (\omega+ i\eta,\mathbf{k}_\parallel )
 = \sum_{\mathbf{k}_\parallel'} \Gamma_{00}^{\chi\chi'\sigma} \mathrm{Im} \mathcal{G}^{\chi'\sigma}_0 (\omega+ i \eta, k_{r,\sigma}' ) 
\end{equation}
with the imaginary part of the clean retarded Green's function
\begin{equation}
 \mathrm{Im} \mathcal{G}^{\sigma}_0 (\omega + i \eta, k_{r,\sigma}' ) 
 = - \pi \delta \left( \omega - \hbar v k_{r,\sigma}' \right) .
\end{equation}
We decompose the sum over momenta into an integral
\begin{equation}
 \sum_{\mathbf{k}'} \to \frac{L^3}{8\pi^3} \int\limits_{-\infty}^\infty d k_r' d q_x \chi' \int\limits_{0}^{\chi' \delta/(2{\ell_5'}^2)}  d k_{\theta}'
\end{equation}
with the integration over $k_\theta'$ depending on the chirality $\chi'$, cf.\ Fig.~\ref{fig:setup}.
Let us for simplicity focus on $\omega = 0$.
We evaluate the self-energy at the momenta $k_z= \chi b_z$ and $k_y \in [-L/(2\ell_5^2), L/(2\ell_5^2)]$ for white noise disorder.
In this simple case, the disorder correlator in momentum space is $K_\mu (\mathbf{q}) = \gamma^2/L^3$, independent of $\mu$ and $\mathbf{q}$.
The disorder correlator in Landau level basis is then
\begin{align}
 \Gamma_{00}^{\chi\chi'\sigma} = \frac{4 \gamma^2}{\sqrt{4 \pi} \sqrt{\ell_5^2+{\ell_5'}^2} }
 e^{ - \frac{(\alpha  \ell - \alpha'\ell_5')^2}{\ell_5^2+{\ell_5'}^2} } ,
\end{align}
which depends on the momentum components $k_{\theta,\sigma}$, $k_{\theta,\sigma}'$.
Since the Green's function depends only	 on the momentum component $k_{r,\sigma}'$, we can compute the integrals over $k_{\theta,\sigma}'$,  $k_{r,\sigma}'$ independently, giving
\begin{equation}
 \frac{1}{L} \sum_{k_{r,\sigma}'} \mathrm{Im} \mathcal{G}^\sigma (\omega+ i\eta , k_{r,\sigma}') = - \frac{1}{2 \hbar v}
\end{equation}
and
\begin{align}
 \frac{1}{L} & \sum_{k_{\theta,\sigma}'} \Gamma_{00}^{\chi\chi'\sigma}
 = \frac{\gamma^2}{\pi^{3/2} \sqrt{\ell_5^2 + {\ell_5'}^2} }
  \chi'\int\limits_0^{\frac{\chi'\delta}{2 {\ell_5'}^2}} d k_{\theta,\sigma}' e^{ - \frac{(\alpha \ell_5 - \alpha'\ell_5')^2}{\ell_5^2+{\ell_5'}^2} } \nonumber \\
 =& \frac{ \gamma^2}{2 \pi {\ell_5'}^2}
  \left[ \mathrm{erf} \left( \frac{\frac{L+\delta}{2} - \sigma\chi \ell_5^2 k_y}{\sqrt{\ell_5^2 + {\ell_5'}^2}} \right)  - \mathrm{erf} \left( \frac{\frac{L}{2} - \sigma\chi \ell_5^2 k_y}{\sqrt{\ell_5^2 + {\ell_5'}^2}} \right)  \right]
\end{align}
Further summing over both surfaces and both surface chiralities $\chi'$ gives
\begin{widetext}
\begin{equation}
 - \mathrm{Im} \Sigma_0^\chi (k_y)
  =  \frac{\gamma^2}{\pi \hbar v {\ell_5'}^2} \left[
  \mathrm{erf} \left( \frac{\frac{L+\delta}{2} + \chi \ell_5^2 k_y}{\sqrt{\ell_5^2 + {\ell_5'}^2}} \right)
 +\mathrm{erf} \left( \frac{\frac{L+\delta}{2} - \chi \ell_5^2 k_y}{\sqrt{\ell_5^2 + {\ell_5'}^2}} \right)
 -\mathrm{erf} \left( \frac{\frac{L}{2} + \chi \ell_5^2 k_y}{\sqrt{\ell_5^2 + {\ell_5'}^2}} \right)
 -\mathrm{erf} \left( \frac{\frac{L}{2} - \chi \ell_5^2 k_y}{\sqrt{\ell_5^2 + {\ell_5'}^2}} \right) \right]
 \label{eq:imaginary_self_energy} ,
\end{equation}
\end{widetext}
independent of the bulk chirality $\chi$.

In Fig.~\ref{fig:analytical_tauv}, we show the imaginary part of the self-energy correction~\eqref{eq:imaginary_self_energy}.
Scattering from states deep in the bulk to the surface is exponentially suppressed due to the exponential localization of the bulk zeroth Landau levels.
Such an exponential suppression would result in almost ballistic states deep in the bulk, up to exponential corrections.
Since this disagrees with our numerical observations, a different scattering mechanism must be more relevant.

As argued by \citeauthor{Pikulin:2016bn}, bulk states in the zeroth Landau levels are localized with the localization length $\ell_5$~\cite{Pikulin:2016bn}.
To scatter between bulk and surface, $N \sim L/\ell_5$ scattering events are necessary to reach the surface from states deep in bulk, which contribute the most to the conductance.
Thus, a random walk from the bulk to one surface requires $N^2 \sim B_5 L^2$ scattering events.
States that are $\Delta L$ apart from the center only need to scatter the smaller distance $L-\Delta L$.
This introduces subleading corrections in $L$, but not in $B_5$.
These subleading orders are especially relevant away from the limit $L \gg \ell_5$, i.e., the regime we investigate in our numerical analysis.
Having a scaling of the transport time $\tau \propto B_5$ is consistent with these numerical observations, as well as $\tau \propto 1 + \alpha L + \beta L^2$ when $\alpha > \beta L$.

\bibliography{ref_new}

\end{document}